\documentclass[prd,a4paper,superscriptaddress,nofootinbib,10pt]{revtex4}
\usepackage{amsfonts}
\usepackage{graphicx}
\usepackage{amssymb}
\usepackage{amsmath}
\usepackage{lscape}
\usepackage{mathrsfs}
\usepackage{textcomp}
\usepackage{epsfig}
\usepackage{color}
\begin{document}

\title{Twisted conformal algebra related to $\kappa$-Minkowski space}
\author{Stjepan Meljanac}
\email{meljanac@irb.hr}
\affiliation{Rudjer Bo\v{s}kovi\'c Institute, Bijeni\v cka c.54, HR-10002 Zagreb, Croatia}
\author{Anna Pacho{\l }}
\email{apachol@unito.it}
\affiliation{Dipartimento di Matematica "Giuseppe Peano", Universita degli Studi di
Torino, Via Carlo Alberto, 10 - 10123 Torino, Italy}
\author{Danijel Pikuti\'c}
\email{dpikutic@irb.hr}
\affiliation{Rudjer Bo\v{s}kovi\'c Institute, Bijeni\v cka c.54, HR-10002 Zagreb, Croatia}

\begin{abstract}
Twisted deformations of the conformal symmetry in the Hopf algebraic framework are constructed.
The first one is obtained by Jordanian twist built up from dilatation and momenta generators. The second is the light-like $\kappa$-deformation of the Poincar\'e algebra extended to the conformal algebra, obtained by twist corresponding to the extended Jordanian r-matrix.
The $\kappa$-Minkowski spacetime is covariant quantum space under both of these deformations. The extension of the conformal algebra by the noncommutative coordinates is presented in two cases, respectively.
The differential realizations for $\kappa$-Minkowski coordinates, as well as their left-right dual counterparts, are also included. 
\end{abstract}

\maketitle
PACS: 11.25.Hf, 16T05, 17B37, 17B81.
\section{Introduction}
The conformal symmetry is considered as the fundamental symmetry of spacetime. Even though it cannot describe massive particles and fields, many high-energy physics theories admit the conformal symmetry. It also includes two fundamental geometries - Poincar\'e and de Sitter - as subcases. The conformal algebra $\mathfrak c$ consists of the Lorentz generators $M_{\mu\nu}$, translations $P_\mu$, dilatations $D$ (which generate scaling transformations) and generators of the special conformal transformations $K_\mu$. 
The metric tensor on the $d$-dimensional spacetime we denote as $g_{\mu\nu}$ (it does not need to be in the diagonal form, it only has to be symmetric
and non-degenerate).
The commutation relations of the conformal algebra $\mathfrak{c}$ 
for $d>2$ (for example for $d=4$ we deal with $\mathfrak{c}=\mathfrak{so}(2,4)$), including the standard Poincar\'{e} ones, are the following:%
\begin{equation}\begin{split} \label{conformalc}
[M_{\mu\nu},M_{\rho\sigma}]  &=
i\left(g_{\mu\sigma}M_{\nu\rho}+g_{\nu\rho} M_{\mu\sigma}
-g_{\mu\rho}M_{\nu\sigma}-g_{\nu\sigma}M_{\mu\rho}\right),  
\\
[M_{\mu\nu},P_\rho]  &= i\left(g_{\nu\rho}P_\mu-g_{\mu\rho}P_\nu\right),  
\\
[D,K_\mu] &= -iK_\mu, \quad [D,P_\mu]=iP_\mu, 
\\
[K_\mu, P_\nu]&=2i(g_{\mu\nu}D - M_{\mu\nu}), \quad  
[K_\mu,M_{\nu\rho}]=i(g_{\mu \nu }K_\rho-g_{\mu\rho}K_\nu),
\\
[K_\mu,K_\nu] &=0,\quad [M_{\mu\nu},D]=0, \quad [P_\mu ,P_\nu] =0
\end{split}\end{equation}

Together with the rise of the interest in the deformations of relativistic symmetries of spacetime \cite{Luk1,Chaichian04} the quantum deformations of the conformal algebra have been investigated already in the 90's \cite{LukNow-q-defConf96}-\cite{Ballesteros-Jordconf}. After the introduction of the $\kappa$-deformed Poincar\'e algebra (with $M_{\mu \nu }$ and $P_{\mu }$ as its generators) with the dimensionfull deformation parameter $\kappa$, the same classical r-matrix as in \cite{Luk1} ($r=\frac{i}{\kappa }M_{0\nu }\wedge P^{\nu }$ with special choice of the basis for which the metric tensor is $g_{00} = 0$) was used in the quantum deformation of Poincar\'e-Weyl algebra \cite{KosMas-kWeyl95}. Also, the deformations of the full $D = 4$ conformal symmetries were introduced \cite{LukKlimek95,LukMinnMozrz95}, corresponding to the standard (i.e. time-like) version of the $\kappa -$deformation. The so-called null-plane (light-cone) deformation of Poincar\'e algebra \cite{Ballesteros-ll} has been extended as well to the deformation of Poincar\'e-Weyl group \cite{KosLukMas-kConfLL} and to conformal group \cite{conf_group1} as well. All of the above mentioned deformations of conformal symmetry were corresponding to the $\kappa$-Minkowski spacetime noncommutativity: $[\hat x^\mu ,\hat x^\nu] =i(a^\mu \hat x^\nu -a^\nu \hat x^\mu )$ in either time-like $a^\mu a_\mu<0$, space-like $a^\mu a_\mu>0$ or light-like $a^\mu a_\mu =0$ cases. The deformation parameter $\kappa$ enters via $a^\mu=\frac1\kappa\tau^\mu$, with $\tau^2\in\{-1,0,1\}$ for the metric tensor with Lorentzian (mostly positive) signature.

Since the Poincar\'{e}-Weyl and conformal algebras contain dilatation $D$ as additional generator, one can also consider another classical r-matrix, $r=\frac i\kappa D \wedge P_0$, as internal one for these algebras and the corresponding quantum deformations. For example in \cite{Ballesteros-Jordconf} 
the so-called Jordanian deformations of the conformal algebra (together with
Anti de Sitter and de Sitter ones) were considered. One should also mention that the same (time-like) Jordanian r-matrix was also used in the twisted deformation of the Poincar\'{e}-Weyl algebra \cite{ABAP09} as well as in the twisting of the inhomogeneous general linear algebra \cite{ABAP10}
together with some applications found in field and gauge theories \cite{DJP, Toward}. A large class of Abelian twists related to twisted statistics in $\kappa$-Minkowski spacetime was considered in \cite{Gov}.

For the conformal algebra the twist type of the deformation was firstly considered in the Moyal-Weyl case \cite{Matlock} related with the Abelian classical r-matrix, $r=i\theta^{\mu\nu} P_\mu\wedge P_\nu$, corresponding to the constant-noncommutativity of the spacetime coordinates $[\hat{x}^\mu,\hat{x}^\nu]=i\theta ^{\mu \nu }$. The $\theta$-deformation of the conformal algebra found some applications in noncommutative field theories, see e.g. \cite{Lizzi-Vitale} and has been extended also to deformation of the superconformal algebra \cite{Super} as well.

Recently, the Jordanian deformations have gained in popularity in the
applications in AdS/CFT correspondence as some of the deformations of the
Yang-Baxter sigma models were shown to preserve the
integrability \cite{Klimcik, Delduc}. The Jordanian deformations related with the classical
r-matrices (satisfying the classical or modified Yang-Baxter equation) were
applied to the AdS part of the correspondence principle \cite{MatYosh}. The $\kappa$-Minkowski spacetime was also considered in this context in \cite{Tongeren}.

Our aim in this paper is to present the quantum deformations of the conformal algebra which are described by the classical r-matrices satisfying classical Yang-Baxter equation (CYBE). For such cases the Drinfeld twists (satisfying the cocycle condition) provide explicitly the star-product in the algebra of spacetime coordinates. We are  interested in the Jordanian $\kappa$-deformation of the conformal symmetry and the light-like $\kappa$-deformation of the Poincar\'e algebra extended to the conformal algebra which will correspond to Jordanian and extended Jordanian r-matrices, respectively. The deformed conformal algebra is considered as Hopf algebra with twisted coproducts and antipodes. Conformal invariance is compatible with the $\kappa$-Minkowski spacetime which constitutes the covariant quantum space under both of these deformations.

The paper is organized as follows.
We start, in Sec.~II, with recalling the basics of the twist deformations and the conditions for noncommutative spacetime covariance with respect to the twisted symmetries. 
In Sec.~III we consider the Jordanian twist \cite{ABAP09, Jord} in the covariant form \cite{Zagreb-jord} providing, for the metric with Lorentzian signature, three kinds (time-, light- and space-like) of deformations of the conformal algebra depending on the type of the vector $a^\mu$. The twisted coalgebra sector is presented together with the corresponding $\kappa$-Minkowski spacetime realization consistent with the Hopf-algebraic actions. 
Later, in Sec.~IV, we investigate another twist \cite{Universal, Light-like, EPJC-1506} 
related by transposition to the so-called extended Jordanian twist \cite{KLM, BP14}. The twist is built from only Poincar\'e generators, therefore it provides the extension of the light-like $\kappa$-deformation of the Poincar\'e algebra to the conformal algebra. For the Poincar\'e subalgebra, the $a^2=0$ deformation reduces to the null-plane deformation of \cite{Ballesteros-ll, BP14}. The realization for $\kappa$-Minkowski coordinates is also presented.
In both cases, in Sec.~III and IV, we include the cross-commutation relations between the conformal algebra generators and the noncommutative coordinates. Also, the so-called left-right dual $\kappa$-Minkowski realizations are constructed from the transposed twists in sections \ref{sectJ} and \ref{sectLL}, respectively. The last section concludes the paper with some remarks.

\section{Twist deformations of the conformal algebra}

The twist deformation framework of spacetime symmetries requires us to deal
with the Hopf algebras instead of Lie algebras corresponding to the given
symmetry. To introduce this notion, we need to extend the Lie algebra
(we are interested in the conformal algebra $\mathfrak{c}$ described by (\ref%
{conformalc})) into the universal enveloping algebra $U\left( \mathfrak{c}%
\right) $ which can be equipped with the Hopf algebra structures on its
generators $L=\{M_{\mu \nu },P_{\mu },D,K_{\mu }\}$ in the following
standard way: 
\begin{align}
\mathrm{coproduct\!:} \qquad  &\Delta\left( L\right) =L\otimes 1+1\otimes L \label{undef-cop}\\
\mathrm{counit\!:} \qquad &\epsilon \left( L\right) =0 \label{undef-cou} \\
\mathrm{antipode\!:} \qquad  &S\left( L\right) =-L. \label{undef-ant} 
\end{align}%
The above maps are then extended to the whole $U\left( \mathfrak{c}\right) $. 
Such undeformed Hopf algebra can be seen as the conformal symmetry of the
usual Minkowski spacetime in the algebraic form given by an Abelian algebra of
coordinate functions $x^\mu \in \mathcal A$, which is itself a subalgebra of undeformed Heisenberg algebra $H$:
\begin{align}
[x^\mu,x^\nu] &=0  \label{ab-alg} \\
[x^\mu,P_\nu] &=i\delta^\mu_\nu \label{heis-xp} \\
[P_\mu,P_\nu] &=0 \label{heis-pp}
\end{align}
The conformal algebra has the following representation:
\begin{equation}\begin{split}
M^{\mu\nu}&=-x^\mu P^\nu + x^\nu P^\mu\\
D&=x\cdot P \\
K^\mu &=2x^\mu (x \cdot P) - x^2 P^\mu = x^\mu D + x_\alpha M^{\alpha\mu}
\end{split}\end{equation}
where $P^\mu=g^{\mu\nu}P_\nu$. In general the compatibility of the spacetime with its symmetry in this
``algebralized'' setting is via the action $\mathcal{H}\otimes \mathcal{A}%
\rightarrow \mathcal{A}$ of the Hopf algebra $\mathcal{H}$ on the spacetime
(module) algebra $\mathcal{A}$ such that%
\begin{equation}
L\triangleright (f\cdot g)=\mu\left[\Delta(L)(\triangleright\otimes \triangleright)(f\otimes g)\right]. \label{genLeibniz}
\end{equation}%
The multiplication in the module algebra $\mu\!: \mathcal A \otimes \mathcal A \rightarrow \mathcal A$ is compatible\footnote{%
It is also common to re-write the condition (\ref{genLeibniz}) as $%
L\triangleright (f\cdot g)=(L_{(1)}\triangleright f)\cdot
(L_{(2)}\triangleright g)$ where Sweedler notation for the coproduct is used $\Delta(L)=L_{(1)}\otimes L_{(2)}$.} with the coproduct in the Hopf algebra $\Delta\!: \mathcal H \rightarrow \mathcal H \otimes \mathcal H$ and $L\left( 1\right) =\epsilon (L)\cdot 1$, $%
1\left( f\right) =f\,\ $for $L\in \mathcal{H}$ and $f\in \mathcal{A}$.


One can easily check that the above condition (\ref{genLeibniz}) is
satisfied for the undeformed spacetime described by Abelian algebra (\ref%
{ab-alg}) and the conformal Hopf algebra (\ref{conformalc}) with (\ref{undef-cop}-\ref{undef-ant}) as its symmetry and the
condition reduces to the usual Leibniz rule: 
\begin{equation}
L\triangleright \left( x^{\mu }\cdot x^{\nu }\right) =\left( L\triangleright
x^{\mu }\right) x^{\nu }+x^{\mu }\left( L\triangleright x^{\nu }\right)
\quad ,\quad L=\{M_{\mu \nu },P_{\mu },D,K_{\mu }\}
\end{equation}%
for any of the generators of the conformal algebra due to (\ref{undef-cop}).

For the deformation we will use the (Drinfeld) twist technique which will
provide the deformation of the universal enveloping algebra of the conformal
algebra $U\left( \mathfrak{c}\right) $ as Hopf algebra $\mathcal{H=}\left(
U\left( \mathfrak{c}\right),\Delta ,\epsilon ,S\right) .$
The twist $\mathcal{F}$ is, in general, an invertible element of $\mathcal{H%
}\otimes \mathcal{H}$ satisfying cocycle and normalization conditions: 
\begin{gather}
(\mathcal{F}\otimes 1)(\Delta \otimes id)\mathcal{F}=(1\otimes \mathcal{F}%
)(id\otimes \Delta )\mathcal{F},  \label{Twcond1} \\
(id\otimes \epsilon )\mathcal{F}=(\epsilon \otimes id)\mathcal{F}=1\otimes 1.
\label{Twcond2}
\end{gather}%
One gets the new Hopf algebra structure $\mathcal{H}^{\mathcal{F}}=\left(U(\mathfrak{c}),
\Delta ^{\mathcal{F}},\epsilon ,S^{\mathcal{F}}\right) $ via modifying the
coproduct and antipode maps in the following way: 
\begin{equation}\begin{split}
\Delta^{\mathcal{F}}(L) &=\mathcal{F}\Delta (L)\mathcal{F}^{-1}
,\quad L\in \mathcal{H}  \\
\epsilon (L) &=0,\quad S^{\mathcal{F}}(L)=\mathrm{f}^{\alpha }S(\mathrm{f%
}_{\alpha })S(L)S(\bar{\mathrm{f}}^{\beta })\bar{\mathrm{f}}_{\beta }.
\label{TwistedUg}
\end{split}\end{equation}%
Here we use the short notation for the twist as: $\mathcal{F}=\mathrm{f}%
^{\alpha }\otimes \mathrm{f}_{\alpha },\quad \mathcal{F}^{-1}=\bar{\mathrm{f}%
}^{\alpha }\otimes \bar{\mathrm{f}}_{\alpha }$.
Both of the twisted deformations considered in this paper will be compatible
with the $\kappa -$Minkowski spacetime with the defining commutation
relations as: 
\begin{equation}
\left[ \hat{x}^{\mu },\hat{x}^{\nu }\right] =i\left( a^{\mu }\hat{x}^{\nu
}-a^{\nu }\hat{x}^{\mu }\right)  \label{cov_kmink}
\end{equation}%
This algebra will constitute the module algebra over the deformed conformal Hopf
algebra, i.e. it is its covariant quantum space.


The cocycle condition (\ref{Twcond1}) for the twist guarantees the
co-associativity of the deformed coproduct $\Delta ^{\mathcal{F}}$ and also
associativity of the corresponding twisted star-product in the twisted
module algebra $\mathcal{A}^{\mathcal{F}}\left( \mathcal{A},\mu _{\star
}\right) $: 
\begin{equation}
f\star g=\mu _{\star }\left( f\otimes g\right) =\mu \circ \mathcal{F}%
^{-1}(\triangleright \otimes \triangleright) (f\otimes g)=(\bar{\mathrm{f}}%
^{\alpha }\triangleright f)\cdot (\bar{\mathrm{f}}_{\alpha }\triangleright g)
\label{tsp}
\end{equation}%
for $f,g\in \mathcal{A}$.
Additionally, to a given twist we can associate the so-called realization of noncommuting
coordinate functions as follows:
\begin{equation}\label{realization}
\hat{x}^{\mu }=\mu \left[  \mathcal{F}^{-1}\left(\triangleright \otimes
1\right) \left( x^{\mu }\otimes 1\right) \right] =(\bar{\mathrm{f}}^{\alpha
}\triangleright x^{\mu })\cdot \bar{\mathrm{f}}_{\alpha }\quad ,\quad x^{\mu
}\in \mathcal{A}
\end{equation}
For the twisted case the compatibility between the deformed coproduct $%
\Delta ^{\mathcal{F}}$ and the $\star $-product in the module algebra is analogous to (%
\ref{genLeibniz}): 
\begin{equation}
L\triangleright \left( \mu _{\star }(f\otimes g)\right) =\mu _{\star }\left(
\Delta ^{\mathcal{F}}\left( L\right) \left(\triangleright \otimes \triangleright\right)
\left( f\otimes g\right) \right)  \label{twgenLeibniz}
\end{equation}%
In the literature this condition is known under twisted covariance and for
example it was investigated in more detail in the context of the
conformal algebra undergoing the Moyal-Weyl deformation with theta-deformed
spacetime \cite{Matlock}. The covariance under twisted symmetry was firstly
proved in \cite{Chaichian04} for the Moyal-Weyl deformation of the Poincar\'{e}
symmetry and theta-spacetime. In the Hopf algebraic framework when the
noncommutative spacetimes are Hopf modules and their deformed symmetry is
the Hopf algebra, the condition of covariance is automatically satisfied
via the requirements \eqref{genLeibniz} and \eqref{twgenLeibniz}.

%


\section{Jordanian deformation of the conformal algebra}\label{sectJ}

We can deform the conformal Hopf algebra $U(\mathfrak c)$ (\ref{conformalc},~\ref{undef-cop},~\ref{undef-cou},~\ref{undef-ant}) with the Jordanian twist \cite{ABAP09,Jord, Zagreb-jord, EPJC-1506}:
\begin{equation}
\mathcal F_J = \exp \left[ i\ln (1-a\cdot P) \otimes D\right]
\label{Twjord}
\end{equation}%
where $a\cdot P=a^\mu P_\mu$. The corresponding classical r-matrix is $r=ia^\mu D\wedge P_\mu$. For the metric with the Lorentzian signature we can distinguish here three cases when vector $a^\mu$ can be either time-like, light-like or space-like, nevertheless the formulae presented below are valid for arbitrary, symmetric
and non-degenerate metric.

For simplicity, we introduce the shortcut notation $Z=1-a\cdot P$.

The algebra relations (\ref{conformalc}) and counits (\ref{undef-cou}) stay undeformed. The deformed coproducts are: 
\begin{align}
\Delta^{\mathcal F_J} (P_\mu) & = P_\mu \otimes 1+Z \otimes P_\mu  \\
\Delta^{\mathcal F_J}(M_{\mu\nu}) & =\Delta (M_{\mu\nu})
-(a_\mu P_\nu -a_\nu P_\mu)Z^{-1} \otimes D \\
\Delta^{\mathcal F_J}(D) & =D\otimes 1+Z^{-1}\otimes D \\
\Delta^{\mathcal F_J}(K_\mu) & =K_{\mu }\otimes
1+Z^{-1}\otimes K_{\mu } \notag \\
& +2[a^{\alpha }M_{\alpha \mu }+a_{\mu }D]Z^{-1}\otimes D \notag \\
& -[2a_{\mu }(a\cdot P)+a^{2}P_{\mu }]Z^{-2}\otimes iD(iD+1)
\end{align}%
And the deformed antipodes: 
\begin{align}
S^{\mathcal{F}_{J}}(P_{\mu })& =-Z^{-1}P_{\mu } \\
S^{\mathcal{F}_{J}}(M_{\mu \nu })& =-M_{\mu \nu }-(a_{\mu }P_{\nu }-a_{\nu
}P_{\mu })D \\
S^{\mathcal{F}_{J}}(D)& =-ZD \\
S^{\mathcal{F}_{J}}(K_{\mu })& =-Z\{K_{\mu }-2[a^{\alpha }M_{\alpha \mu
}+a_{\mu }D]D \notag \\
& ~\qquad  +[2a_{\mu }(a\cdot P)+a^{2}P_{\mu }]iD(iD+1)\}
\end{align}%
The corresponding covariant quantum spacetime is the $\kappa $-Minkowski one
(\ref{cov_kmink}) 
with the realization for coordinates given via (\ref{realization}) as:%
\begin{equation}\label{hatxJ}
\hat{x}_{J}^{\mu }=\mu \left[\mathcal F^{-1}_J(\triangleright\otimes1)(x^\mu\otimes1) \right] =x^{\mu }-a^{\mu }D = x^\mu - a^\mu (x\cdot P)
\end{equation}%
It is called right covariant realization \cite{x1110, KJMS}, and commutators with generators of the conformal algebra are:
\begin{equation}\begin{split}
[P^\mu, \hat x_J^\nu] &=-i(g^{\mu\nu}-a^\nu P^\mu) \\
[D, \hat x_J^\mu] &=-ix^\mu \\
[M^{\mu\nu}, \hat x_J^\lambda] &=i(x^\mu g^{\nu\lambda} - x^\nu g^{\mu\lambda}) \\
[K^\mu, \hat x_J^\nu] &=i(2x^2 g^{\mu\nu} - x^\mu x^\nu - a^\nu K^\mu)
\end{split}\end{equation}
where
\begin{equation}
x^\mu = \hat x_J^\mu + a^\mu D
\end{equation}
Note that the commutators are closed in the conformal algebra and noncommutative coordinates $\hat x_J^\mu$. 

The spacetime algebra (\ref{cov_kmink}) obtained via (\ref{tsp}) is invariant under the twisted conformal transformations
which can be seen from action of the conformal symmetry generators on the
algebra of functions of $\kappa$-Minkowski coordinates, i.e. via the
compatibility condition (\ref{twgenLeibniz}). One can check that indeed: 
\begin{equation}\label{inv}
L\triangleright
\left[\mu\circ\mathcal F_J^{-1}(\triangleright\otimes\triangleright)(x^\mu\otimes x^\nu - x^\nu\otimes x^\mu) \right]
=L\triangleright \left[ i\left( a^\mu x^\nu -a^\nu x^\mu \right) \right]
\end{equation}%
the twisted case of the Leibniz rule is satisfied for any of the
generators $L=\{M_{\mu \nu },P_{\mu },D,K_{\mu }\}$.


Transposed twist $\tilde{\mathcal F_J}=\tau_0 \mathcal F_J \tau_0$ is obtained from $\mathcal F_J$ by interchanging left and right side of tensor product (i.e. $\tau_0(a\otimes b)=b\otimes a$), and it is also a Drinfeld twist satisfying cocycle \eqref{Twcond1} and normalization condition \eqref{Twcond2}. A set of left-right dual generators of $\kappa$-Minkowski space can be obtained from transposed twist:
\begin{equation}
\hat y_J^\mu =\mu \left[\tilde{\mathcal F_J}^{-1}(\triangleright\otimes1)(x^\mu\otimes1) \right] = x^\mu(1-a\cdot P)
\end{equation}
Generators $\hat y_J^\mu$ satisfy $\kappa$-Minkowski algebra with $a_\mu\rightarrow-a_\mu$:
\begin{equation}
[\hat y_J^\mu, \hat y_J^\nu] = -i(a^\mu \hat y_J^\nu - a^\nu \hat y_J^\mu)
\end{equation}
and they commute with generators $\hat x_J^\mu$:
\begin{equation}
[\hat x_J^\mu, \hat y_J^\nu] = 0
\end{equation}


\section{Light-like $\protect\kappa -$deformation of Poincar\'e algebra
extended to the conformal algebra}\label{sectLL}

For the purpose of this section we consider the metric tensor with Lorentzian signature and use mostly positive sign convention, i.e. $g_{\mu\nu} = diag(-,+,+,...,+)$.
We deform the conformal Hopf algebra $U(\mathfrak{c})$ (\ref{conformalc},~\ref{undef-cop},~\ref{undef-cou},~\ref{undef-ant}) with the twist $\mathcal F_{LL}$ leading to light-like $\kappa$-deformation of Poincar\'e algebra \cite{BP14,Toward, Universal, EPJC-1506} (which is related to extended Jordanian twist \cite{KLM},\cite{Tolstoy})\footnote{The explicit relation between the twist (\ref{Twlight}) and the standard extended Jordanian twist corresponding to light-like case (up to the transposition) is presented in detail in sect. VIII B in ref. \cite{EPJC-1506} }:
\begin{equation}
\mathcal{F}_{LL}=\exp \left[ -ia_\alpha P_\beta \frac{\ln(1+a\cdot P)}{a\cdot P}\otimes M^{\alpha\beta}\right] 
\label{Twlight}
\end{equation}%
Above twist satisfies the cocycle condition (\ref{Twcond1}) \cite{EPJC-1506} with the light-like vector $a^\mu$ \cite{Universal, EPJC-1506} and the classical r-matrix is $r=a^\mu M_{\mu\nu}\wedge P^\nu$. We also introduce the following notation: 
\begin{equation}
\tilde Z=\frac{1}{1+a\cdot P},\qquad m_{\mu }=a^{\alpha }M_{\alpha \mu }
\end{equation}%

Coproducts: 
\begin{align}
\Delta ^{\mathcal{F}_{LL}}\left( P_{\mu }\right) & =\Delta \left( P_{\mu
}\right) +\left[ P_{\mu }a^{\alpha }-a_{\mu }\left( P^{\alpha }+\frac{1}{2}%
a^{\alpha }P^{2}\right) \tilde Z\right] \otimes P_{\alpha } \\
\Delta ^{\mathcal{F}_{LL}}\left( M_{\mu \nu }\right) & =\Delta \left( M_{\mu
\nu }\right) +(\delta _{\mu }^{\alpha }a_{\nu }-\delta _{\nu }^{\alpha
}a_{\mu })\left( P^{\beta }+\frac{1}{2}a^{\beta }P^{2}\right) \tilde Z\otimes
M_{\alpha \beta } \\
\Delta ^{\mathcal{F}_{LL}}\left( m_{\alpha }\right) & =\Delta \left(
m_{\alpha }\right) +a_{\mu }\left( P^{\alpha }+\frac{1}{2}a^{\alpha
}P^{2}\right) \tilde Z\otimes m_{\alpha } \\
\Delta ^{\mathcal{F}_{LL}}\left( D\right) & =\Delta \left( D\right)
-P_{\alpha }\tilde Z\otimes m^{\alpha } \\
\Delta ^{\mathcal{F}_{LL}}\left( K_{\mu }\right) & =K_{\mu }\otimes 1+\left[
\delta _{\mu }^{\alpha }+P_{\mu }a^{\alpha }-a_{\mu }\left( P^{\alpha }+%
\frac{1}{2}a^{\alpha }P^{2}\right) \tilde Z\right] \otimes K_{\alpha } \notag \\
& +\left\{ \left[ 2\left( a_{\mu }(iD+\tilde Z)-im_{\mu }\right) P^{\alpha }+i{%
M_{\mu }}^{\alpha }\right] -2D g _{\mu \alpha }\right\} \otimes m_{\alpha }
\notag \\
& +\left[ iP_{\mu } g^{\alpha \beta }-2i(\delta _{\mu }^{\alpha }-a_{\mu
}P^{\alpha }\tilde Z)P^{\beta }\tilde Z\right] \otimes m_{\alpha }m_{\beta }
\end{align}%
Antipodes: 
\begin{align}
S^{\mathcal{F}_{LL}}(P_{\mu })& =-\left[ P_{\mu }+a_{\mu }\left( P^{\alpha }+%
\frac{1}{2}a^{\alpha }P^{2}\right) P_{\alpha }\right] \tilde Z \\
S^{\mathcal{F}_{LL}}(M_{\mu \nu })& =-M_{\mu \nu }-(\delta _{\mu }^{\alpha
}a_{\nu }-\delta _{\nu }^{\alpha }a_{\mu })\left( P_{\beta }+\frac{1}{2}%
a_{\beta }P^{2}\right) M_{\alpha \beta } \\
S^{\mathcal{F}_{LL}}(m_{\mu })& =-m_{\mu }-a_{\mu }\left( P^{\alpha }+\frac{1%
}{2}a^{\alpha }P^{2}\right) m_{\alpha } \\
S^{\mathcal{F}_{LL}}(D)& =-\tilde Z^{-1}D \end{align}%
\begin{align}
S^{\mathcal{F}_{LL}}(K_{\mu })& =-\left[ \delta _{\mu }^{\gamma }+P^{\gamma
}a^{\mu }-a^{\gamma }\left( P_{\mu }+\frac{1}{2}a_{\mu }P^{2}\right) \tilde Z\right]
\times\notag   \\
& \times \left\{ K_{\gamma }+\left[ 2\left( a_{\gamma }(iD+\tilde Z)-im_{\gamma }\right)
P^{\alpha }+i{M_{\gamma }}^{\alpha }\right] S(m_{\alpha })-2DS(m_{\gamma }) \right.
\notag \\
& ~\quad \left. +iP_{\gamma }S(m^{2})-2i\left( \delta _{\gamma }^{\alpha }-a_{\gamma
}P^{\alpha }\tilde Z\right) P^{\beta }\tilde ZS(m_{\alpha }m_{\beta }) \right\}
\end{align}%
In this case, the realization (\ref{realization}) is given as: 
\begin{equation}\label{hatxLL}
\hat x_{LL}^\mu =
\mu \left[\mathcal F^{-1}_{LL}(\triangleright\otimes1)(x^\mu\otimes1) \right] =
x^\mu +a_\alpha M^{\alpha\mu} = x^\mu(1+a\cdot P) - (a\cdot x) P^\mu
\end{equation}
It corresponds to the natural realization of $\kappa$-Minkowski space \cite{x1110, KJMS}. Commutators with generators of the conformal algebra are:
\begin{equation}\begin{split}\label{comm-xLLc}
[P^\mu, \hat x_{LL}^\nu] &=-i \left[g^{\mu\nu}(1+a\cdot P) - a^\mu P^\nu \right] \\
[D, \hat x_{LL}^\mu] &=-ix^\mu \\ 
[M^{\mu\nu}, \hat x_{LL}^\lambda] &=i(\hat x_{LL}^\mu g^{\nu\lambda} - \hat x_{LL}^\nu g^{\mu\lambda}-a^\mu M^{\nu\lambda} + a^\nu M^{\mu\lambda}) \\
[K^\mu, \hat x_{LL}^\nu] &=i(2x^2 g^{\mu\nu} - x^\mu x^\nu + a^\mu K^\nu - g^{\mu\nu}(a\cdot K))
\end{split}\end{equation}
where
\begin{equation}
x^\mu = \hat x_{LL}^\mu -a_\alpha M^{\alpha\mu}
\end{equation}
Note that, like in Jordanian case, above commutators \eqref{comm-xLLc} are also closed in the conformal algebra and noncommutative coordinates $\hat x_{LL}^\mu$.

Let us also comment on the fact that even though the above twist (\ref{Twlight}) is written in a covariant form (valid for the $a^\mu$ as time-, light- and space-like vector) it satisfies the cocycle condition (\ref{Twcond1}) only for the light-like case, $a^2=0$~\cite{Universal, EPJC-1506}. Therefore, only in this case it corresponds to an associative star-product (\ref{tsp}) of $\kappa $-Minkowski coordinates (\ref{cov_kmink}). The two remaining cases (time- and space-like) lead to a  deformations of $\kappa$-Snyder type with non-associative star product.

Again, one can easily check that the noncommutative spacetime \eqref{cov_kmink} is invariant under this twisted conformal symmetry via analogous condition as \eqref{inv} in section \ref{sectJ}:
\begin{equation}
L\triangleright
\left[\mu\circ\mathcal F_{LL}^{-1}(\triangleright\otimes\triangleright)(x^\mu\otimes x^\nu - x^\nu\otimes x^\mu) \right]
=L\triangleright \left[ i\left( a^\mu x^\nu -a^\nu x^\mu \right) \right]
\end{equation}%


Transposed twist $\tilde{\mathcal F}_{LL}=\tau_0 \mathcal F_{LL} \tau_0$ is obtained from $\mathcal F_{LL}$ by interchanging left and right side of tensor product, and it is also a Drinfeld twist satisfying cocycle \eqref{Twcond1} and normalization condition \eqref{Twcond2}. A set of left-right dual generators of $\kappa$-Minkowski space can be obtained from transposed twist:
\begin{equation}
\hat y_{LL}^\mu =\mu \left[\tilde{\mathcal F}_{LL}^{-1}(\triangleright\otimes1)(x^\mu\otimes1) \right]
= x^\mu + (a\cdot x) P^\mu - a^\mu \left(D + \frac{a\cdot x}2P^2 \right)\tilde Z
\end{equation}
Generators $\hat y_{LL}^\mu$ satisfy $\kappa$-Minkowski algebra with $a_\mu\rightarrow-a_\mu$:
\begin{equation}
[\hat y_{LL}^\mu, \hat y_{LL}^\nu] = -i(a^\mu \hat y_{LL}^\nu - a^\nu \hat y_{LL}^\mu)
\end{equation}
and they commute with generators $\hat x_{LL}^\mu$:
\begin{equation}
[\hat x_{LL}^\mu, \hat y_{LL}^\nu] = 0
\end{equation}
Realizations $\hat y_J^\mu$ and $\hat y_{LL}^\mu$ cannot be expressed in terms of $x^\mu$ and generators of conformal algebra (whereas realizations $\hat x_J^\mu$ and $\hat x_{LL}^\mu$ are expressed in terms of these generators).


Note that $\hat x^\mu_J$ (eq.\eqref{hatxJ}) and $\hat x^\mu_{LL}$ (eq.\eqref{hatxLL}) are different realizations of $\kappa$-Minkowski space $\hat x^\mu_J \ne \hat x^\mu_{LL}$, related by similarity transformation. There is also another point of view, so that, for $a^2=0$, $\hat x^\mu_J$ and $\hat x^\mu_{LL}$ can be identified, but generators $x^\mu$ and generators of conformal algebra have different realizations in two cases (sections \ref{sectJ} and \ref{sectLL}), related by similarity transformation. In this case, let us denote as $(x_J^\mu, P_J^\mu)$ and $(x_{LL}^\mu, P_{LL}^\mu)$ two pairs of commutative coordinates and momenta, each satisfying undeformed Heisenberg algebra \eqref{ab-alg}-\eqref{heis-pp}, which are related by similarity transformation: 
\begin{align}
P_J^\mu &= \left(P_{LL}^\mu + \frac{a^\mu}2P_{LL}^2 \right)\tilde Z_{LL} \\
P_{LL}^\mu &= \left(P_J^\mu - \frac{a^\mu}2P_J^2 \right)Z_J^{-1} \end{align}%
\begin{align}
x_J^\mu &= \left[x_{LL}^\mu +a^\mu (x_{LL}\cdot P_{LL})\right] \tilde Z_{LL}^{-1} - (a\cdot x_{LL}) \left(P_{LL}^\mu + \frac{a^\mu}2P_{LL}^2 \right) \\
x_{LL}^\mu &= \left[x_J^\mu -a^\mu (x_J\cdot P_J)\right] Z_J + (a\cdot x_J) \left(P_J^\mu - \frac{a^\mu}2P_J^2 \right) 
\end{align}
where
\begin{equation}
Z_J \equiv 1-a\cdot P_J = \frac1{1+a\cdot P_{LL}} \equiv \tilde Z_{LL}.
\end{equation}
Hence, $\hat x^\mu = x_J^\mu - a^\mu (x_J\cdot P_J) = x_{LL}^\mu(1+a\cdot P_{LL}) - (a\cdot x_{LL}) P_{LL}^\mu$ and two sets of conformal generators, $\left\{P_J^\mu, M_J^{\mu\nu}, D_J, K_J^\mu \right\}$ and  $\left\{P_{LL}^\mu, M_{LL}^{\mu\nu}, D_{LL}, K_{LL}^\mu \right\}$, are related by similarity transformation.


\section{Concluding remarks}

We have presented the two different $\kappa$-deformations of the conformal symmetry within the Drinfeld twist framework. Both twists provide the $\kappa$-Minkowski star product, therefore the $\kappa$-Minkowski spacetime stays covariant under the twisted conformal symmetries. Thanks to the twist, we are also able to obtain the differential realization for the noncommutative coordinates.
The extension of the conformal algebra by the noncommutative coordinates is also presented and it includes the deformed phase space (deformed Heisenberg algebra) as subalgebra. 
For alternative point of view, where the phase space stays undeformed but the realizations of the conformal algebra generators are modified, see e.g. \cite{Ghosh}. 
Additionally, we have constructed, from transposed twists, another set of realizations satisfying the $\kappa$-Minkowski relations (with $a^\mu\rightarrow  - a^\mu$).  Both of the deformations presented in this paper (Jordanian and extended Jordanian) provide the so-called 
triangular deformation as the corresponding classical r-matrices satisfy the classical Yang-Baxter equation. 
Interestingly, the Jordanian and extended Jordanian deformations can be generated by other (than already mentioned) classical r-matrices. One can notice that the form of the conformal algebra (%
\ref{conformalc}) does not change if we exchange the generators (see also
similar comment in \cite{LukMinnMozrz95}) in the following way: 
\begin{equation}
P_\mu \rightarrow K_\mu, \qquad K_\mu \rightarrow P_\mu , \qquad
D \rightarrow -D, \qquad \kappa \rightarrow \frac1{\tilde\kappa}
\label{exchange}
\end{equation}
This allows us, to distinguish yet another classical $r$-matrix for the
conformal algebra (besides for example the one investigated in Sec.~\ref{sectJ} for the Jordanian case $r=ia^{\mu }D\wedge
P_{\mu }$), i.e.: 
$r=-i\tilde{a}^{\mu }D\wedge K_{\mu }$
with a new deformation parameter $\tilde{\kappa}$ and  
$\tilde a^\mu=\tilde\kappa^2 a^\mu$.   
 Such r-matrix is satisfying the classical Yang-Baxter equation and the classical limit is obtained for $\tilde{\kappa}%
\rightarrow 0$ (which corresponds to $\kappa \rightarrow \infty $). The new Jordanian twist (\ref{Twjord}) with (\ref{exchange}) for any $\tilde{a}^{\mu }$ will
satisfy the cocycle condition (\ref{Twcond1}) as well. Formal expressions for the twisted deformation of coproducts and antipodes in $\left( U\left( \mathfrak{c}\right) ,\Delta ^{%
\mathcal{F}},\epsilon ,S^{\mathcal{F}}\right) $ (as twisted conformal Hopf
algebra) generated by this r-matrix will stay the same up to (\ref{exchange}).

One way for interpreting the exchange in the deformation parameter $\kappa\rightarrow\frac{1}{\tilde{\kappa}}$ (related with $a_\mu \rightarrow \tilde a_\mu$ as above) could be the following. Instead of considering the minimal length, as it happens when introducing the noncommutative coordinates $\hat{x}^\mu$, we should consider the minimal momentum and introduce the noncommutative momenta $\hat{p}^\mu$. This way the $\tilde{\kappa}$-deformation would appear in the momentum space $[\hat{p}^\mu,\hat{p}^\nu]=i(\tilde{a}^\mu\hat{p}^\nu-\tilde{a}^\nu\hat{p}^\mu)$
instead of (\ref{cov_kmink}). Other physical consequences of such exchange are still an open issue.

Nevertheless, the deformations of the conformal symmetry introduced in this paper can be of interest in many physical applications. For example, the Jordanian deformations are also appearing in the context of AdS/CFT correspondence \cite{MatYosh, Tongeren}, therefore the corresponding deformations of the conformal field theory part in the twisted framework could be of interest as well.
Another point to consider would be, for example, the extension of the deformations introduced in this paper to the supersymmetric case, as it was already considered for the Moyal-Weyl deformation of the conformal superalgebra \cite{Super}. Additionally extending the presented framework into the Hopf algebroid language \cite{Zagreb-jord} would allow to introduce yet another example for the twisted deformation of Hopf algebroids as well.
Also, the deformations of the conformal symmetry presented here could be considered as a starting point in the study on deformed (noncommutative) cosmology. Recently a short review on models of the inflating Universe based on conformal symmetry was presented \cite{Rubtsov}. The straightforward way to make them noncommutative would be to introduce the star-product (\ref{tsp}) related with the twists in the conformally invariant actions corresponding to different models. This way one could investigate, for example, if introducing the deformation parameter (as quantum gravity scale) would have any influence on the scale-invariance of the power spectrum of the scalar perturbations.

 Our results, however, provide only the starting point for such
investigations.

\section*{Acknowledgements}

A. P. acknowledges the funding from the European
Union's Seventh Framework program for research and
innovation under the Marie Sk{\l}odowska-Curie Grant No. 609402–2020 researchers: Train to Move
(T2M). Part of this work was supported by National Science Center Project No. 2014/13/B/ST2/04043. 
The work by S.M. and D.P. has been fully supported by Croatian Science Foundation under the Project No. IP-2014-09-9582.

\end{document}